\documentclass[preprint,preprintnumbers,amsmath,amssymb]{revtex4}

\usepackage{graphicx}
\usepackage{dcolumn}
\usepackage{bm}

\begin{document}

\title{Coupled Modulated Bilayers}

\author{Y. Hirose and S. Komura}
\affiliation{%
Department of Chemistry,
Graduate School of Science and Engineering,
Tokyo Metropolitan University, Tokyo 192-0397, Japan
}%
\author{D. Andelman}
\affiliation{%
Raymond and Beverly Sackler School of Physics and Astronomy,
Tel Aviv University, Ramat Aviv, Tel Aviv 69978, Israel
}%

\date{August 3, 2009}

\begin{abstract}

We propose a model addressing the coupling mechanism between two
spatially modulated monolayers.
We obtain the mean-field phase diagrams of coupled bilayers when
the two monolayers have the same preferred modulation wavelength.
Various combinations of the monolayer modulated phases are obtained and
their relative stability is calculated.
Due to the coupling, a spatial modulation in one of the monolayers
induces a similar periodic structure in the second one.
We have also performed numerical simulations for the case when the 
two monolayers have different modulation wavelengths.
Complex patterns may arise from the frustration between the
two incommensurate but annealed structures.
\end{abstract}

\maketitle

\section{Introduction}
\label{introduction}

Quite a number of physical, chemical, and biological systems manifest
some type of modulation in their spatial ordering~\cite{SA,AR1,AR2}.
Such structures are stripes and bubbles in two-dimensional (2D)
systems, or lamellae, hexagonally packed cylinders, and cubic
arrays of spheres in three-dimensional (3D) cases as well as more
complex structures such as gyroids.
Examples of such systems include ferromagnetic
layers~\cite{DBBEGHJOPPT}, magnetic garnet films~\cite{GD},
ferrofluids~\cite{AR1,AR2,Rosensweig}, dipolar Langmuir
films~\cite{ABJ}, rippled phases in lipid bilayers~\cite{Sackmann1},
and block copolymers~\cite{Hamley,Fredrickson}.
Modulated phases may also occur in systems described by two
(or more) coupled order parameters, each favoring a different
equilibrium state~\cite{LA}.
The observed spatial patterns exhibit striking similarity even for
systems that are very different in their nature.
It is generally understood that the modulated structures are
formed spontaneously due to the competition between short-
and long-range interactions.

In the case of 2D ferromagnetic layers, for example, the
short-range interaction arises from magnetic domain wall energy,
while the long-range interaction is due
to magnetic dipole-dipole interaction which induces a demagnetizing
field~\cite{AR1,AR2}.
Adding both contributions and minimizing the total free energy
with respect to the wavenumber $q$, one obtains the most
stable mode $q^{\ast} \neq 0$.
This description is valid in the weak segregation limit (close to a
critical point), where the equilibrium domain size is given by
$d^{\ast}=2\pi/q^{\ast}$.
In general, this quantity depends on temperature and/or other external
fields.

In this paper, we shall consider two modulated
monolayers that are jointly coupled.
Our motivation is related to recent experiments by Collins and
Keller~\cite{CK} who investigated Montal-Mueller planar bilayer
membranes~\cite{MM} composed of lipids and cholesterol.
In this technique, a bilayer is constructed by preparing separately two 
independent monolayers and then combining them into one joint bilayer
across a hole at the air/water interface.
The experiments addressed specifically the question of liquid domains 
in the two leaflets, and the mutual influence of the monolayers in 
terms of their domain phases.
In the experiment, asymmetric bilayers are prepared in such a way 
that one leaflet's composition would phase-separate in a symmetric 
bilayer and the other's would not.
In some cases, one leaflet may induce phase separation in the other
leaflet, whereas in other cases, the second leaflet suppresses domain
formation in the original leaflet.
These results imply that the two leaflet coupling is important ingredient
in determining the bilayer phase state.

Motivated by these experiments,
the coupled bilayer system was investigated theoretically.
The coupling mechanism arises through
interactions between lipid tails across the bilayer midplane, and
the phase behavior of such a bilayer membrane was computed
using either regular solution theory~\cite{WLM} or Landau
theory~\cite{PS}.
The theoretical results are in accord with several of the experimental
observations.
It should be noted that all previous models dealt with the
coupling between two {\it macro--phase} separated leaflets, while it is
also of interest to investigate the coupling between
two {\it micro--phase} separated (modulated) leaflets.
Furthermore, one might also consider the interplay between a
macro-- and a micro--phase separation.

In the present work, we suggest a model describing the
coupling between two modulated systems, and, in particular,
we analyze the influence of this coupling on the phase behavior
of two coupled 2D monolayers.
When the two monolayers have the same preferred periodicity of modulation,
we obtain the mean-field phase diagrams which exhibit various
combinations of micro--phase separated structures.
In some cases, the periodic structure in one of the monolayers will
induce a modulation in the other monolayer.
Interesting situations take place when the two monolayers have
different preferred wavelengths of modulation.
Here the frustrations between the two competing modulated structures
need to be optimized.
These structures and their dynamical behavior are examined using
numerical simulations.
Although there has been so far no experiment which directly corresponds
to the proposed model, our predictions may be verified, for example,
by constructing Montal-Mueller bilayers~\cite{MM} out of two lipid
monolayers that exhibit a striped phase near the miscibility critical
point~\cite{KPHM,KM}.

In the next section, we present a phenomenological model
describing the coupling between two modulated lipid monolayers.
In Sec.~\ref{identical}, we discuss the case when the two monolayers
have the same preferred wavelength of modulation.
Monolayers having different preferred
wavelengths will be considered in Sec.~\ref{different}, and some
related situations are further discussed in Sec.~\ref{discussion}.
Although we limit our present analysis to 2D systems, the
suggested model can be generalized to 3D systems as well.

\section{Model}
\label{model}

\begin{figure}[t!]
\begin{center}
\includegraphics[scale=0.9]{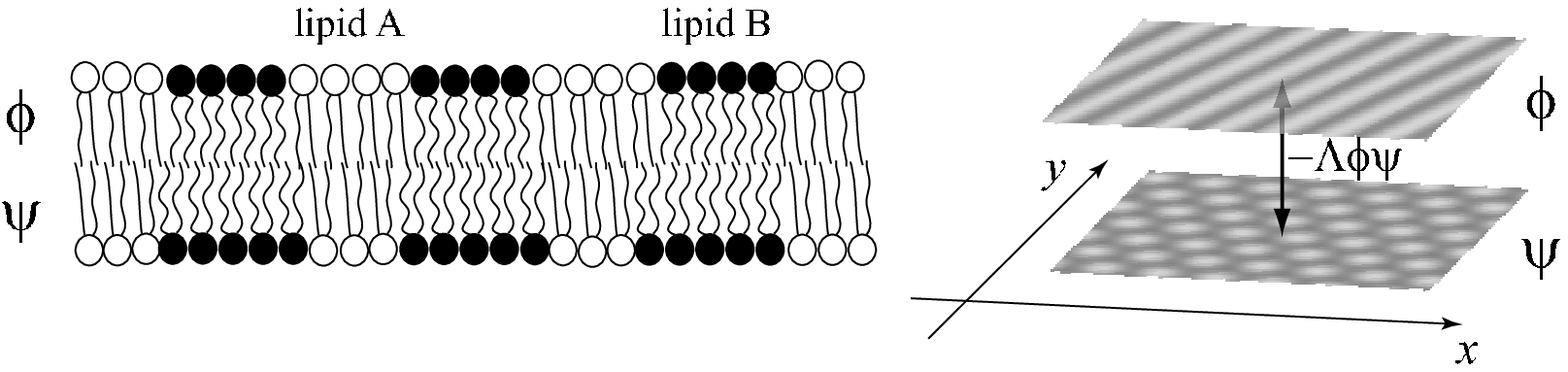}
\end{center}
\caption{\textsf{
Schematic illustration of two coupled modulated monolayers forming
a bilayer membrane.
Each monolayer is composed of a binary A/B lipid mixture, which can have a
spatial modulation.
The relative composition of the two lipids in the upper and the
lower leaflets are defined by $\phi$ and $\psi$, respectively.
In general, the average composition in the two monolayers can be
different.
The lipid tails interact across the bilayer midplane.
The phenomenological coupling term between these two variables are assumed
to be bilinear of the form ${-}\Lambda \phi \psi$ in the free energy of
Eq.~(\ref{freeenergy}).
}}
\label{fig1}
\end{figure}

In order to illustrate the coupling effect between two modulated
systems, we imagine a pair of lipid monolayers forming a coupled
bilayer. Each of the monolayer can
undergo separately a micro--phase separation.
As shown in Fig.~\ref{fig1}, we assume that each monolayer is a
mixture of two lipid species, say lipid A and lipid B.
Their area fractions are defined by $\phi_{\rm A}(\mathbf{r})$ and
$\phi_{\rm B}(\mathbf{r})$, where $\mathbf{r}=(x,y)$ is the 2D
positional vector.
By assuming that the monolayer is incompressible,
$\phi_{\rm A}(\mathbf{r})+\phi_{\rm B}(\mathbf{r})=1$, the monolayer
composition can be characterized by a single order parameter defined
by the relative A/B composition
$\phi(\mathbf{r})=\phi_{\rm A}(\mathbf{r})-\phi_{\rm B}(\mathbf{r})$.
Let us denote this local order parameter of the upper and
lower monolayers by $\phi(\mathbf{r})$ and $\psi(\mathbf{r})$,
respectively.
The coarse-grained free-energy functional for the coupled
modulated bilayer is written as:
\begin{align}
F[\phi, \psi ] & =F_{\rm u}[\phi]+F_{\rm \ell}[\psi]-
\Lambda \int {\rm d}\mathbf{r}\, \phi\psi\nonumber\\
&=\int {\rm d}\mathbf{r}
\biggl[2 (\nabla^2 \phi)^2 - 2 (\nabla \phi)^2
+ \frac{\tau}{2}\phi^2 + \frac{1}{4}\phi^4 - \mu_{\phi}\phi
\nonumber \\
& + 2 D(\nabla^2 \psi)^2 - 2 C(\nabla \psi)^2
+ \frac{\tau}{2}\psi^2 + \frac{1}{4}\psi^4 - \mu_{\psi}\psi
-\Lambda \phi \psi \biggr].
\label{freeenergy}
\end{align}
This is a modified Ginzburg-Landau free energy expanded
in powers of the order parameters $\phi$ and $\psi$ and their derivatives.
The $F_{\rm u}[\phi]$ free energy has five terms depending only on 
$\phi$ and its derivatives.
It describes the upper monolayer and its possible modulations, while
the coefficients of the Laplacian squared, the gradient squared and the
$\phi^4$ terms are taken to be numbers, for simplicity.
Similarly, $F_{\rm \ell}[\psi]$ describing the lower monolayer
contains the next five terms that are only functions of $\psi$ and 
its derivatives.
The last term represents the coupling between the two leaflets as
will be explained later.
The coefficients of the two gradient squared terms are both negative
($C>0$) favoring spatial modulations, whereas the coefficients of the Laplacian
squared terms are positive ($D>0$) to have a  stable modulation at finite
wavenumbers.
The $\phi^2$, $\phi^4$, $\psi^2$ and $\psi^4$
terms in $F$ are the usual Landau
expansion terms with $\tau=(T-T_{\rm c})/T_{\rm c}$ being the reduced
temperature ($T_{\rm c}$ is the critical temperature).
For simplicity, the two leaflets are taken to have the same
critical temperature $T_{\rm c}$ (and hence the same $\tau$).
Finally, the linear term coefficients, $\mu_{\phi}$ and $\mu_{\psi}$,
are the chemical potentials which regulate the average values of $\phi$
and $\psi$, respectively.

In the absence of the coupling term ($\Lambda=0$),
each of the two leaflets can have its own modulated phase.
Free energy functionals such as  $F_{\rm u}$ have been used successfully
in the past to describe a variety of modulated systems: magnetic garnet 
films~\cite{GD}, Langmuir films~\cite{ABJ}, diblock 
copolymers~\cite{Leibler,Hamley}, and amphiphilic systems~\cite{GS}. 
Furthermore, interfacial properties between different coexisting phases 
have been investigated using a similar model~\cite{NAS,VGA,VGNAS}.
In the above expression for the free energy $F$,
the $\phi$-leaflet has a dominant
wavenumber $q^{\ast}_{\phi}=1/\sqrt{2}$, and so has the
$\psi$-leaflet with $q^{\ast}_{\psi}=\sqrt{C/2D}$.
The modulation wavenumbers and amplitudes of the two monolayers
coincide when $D=C=1$ and the average compositions are the same.

Next we address the physical origin of the coupling term 
${-}\Lambda \phi \psi$.
We first note that this quadratic term is invariant under the exchange
of $\phi \leftrightarrow \psi$.
When $\Lambda>0$, this term can be obtained from a $(\phi-\psi)^2$
term~\cite{WLM,PS}, which represents a local energy penalty when the
upper and lower monolayers have different compositions.
In the case of mixed lipid bilayers, such a coupling may result from
the conformational confinement of the lipid chains, and hence would
have entropic origin~\cite{WLM}.
By estimating the degree of the lipid chain interdigitation, the
magnitude of the coupling parameter $\Lambda$ was recently
estimated by May~\cite{May}.
In general, the coupling constant $\Lambda$ can also be negative
depending on the specific coupling mechanism~\cite{May}.
However, it will be explained later that the phase diagram for
$\Lambda<0$ can easily be obtained from the $\Lambda>0$ one.
Hence, it is sufficient to consider only the $\Lambda>0$ case
without loss of generality.
Although the microscopic origin of the coupling may differ between
systems, we will regard $\Lambda$ as a phenomenological parameter
and investigate its role on the structure, phase behavior and dynamics
of coupled modulated bilayers.

\begin{figure}[t!]
\begin{center}
\includegraphics[scale=0.8]{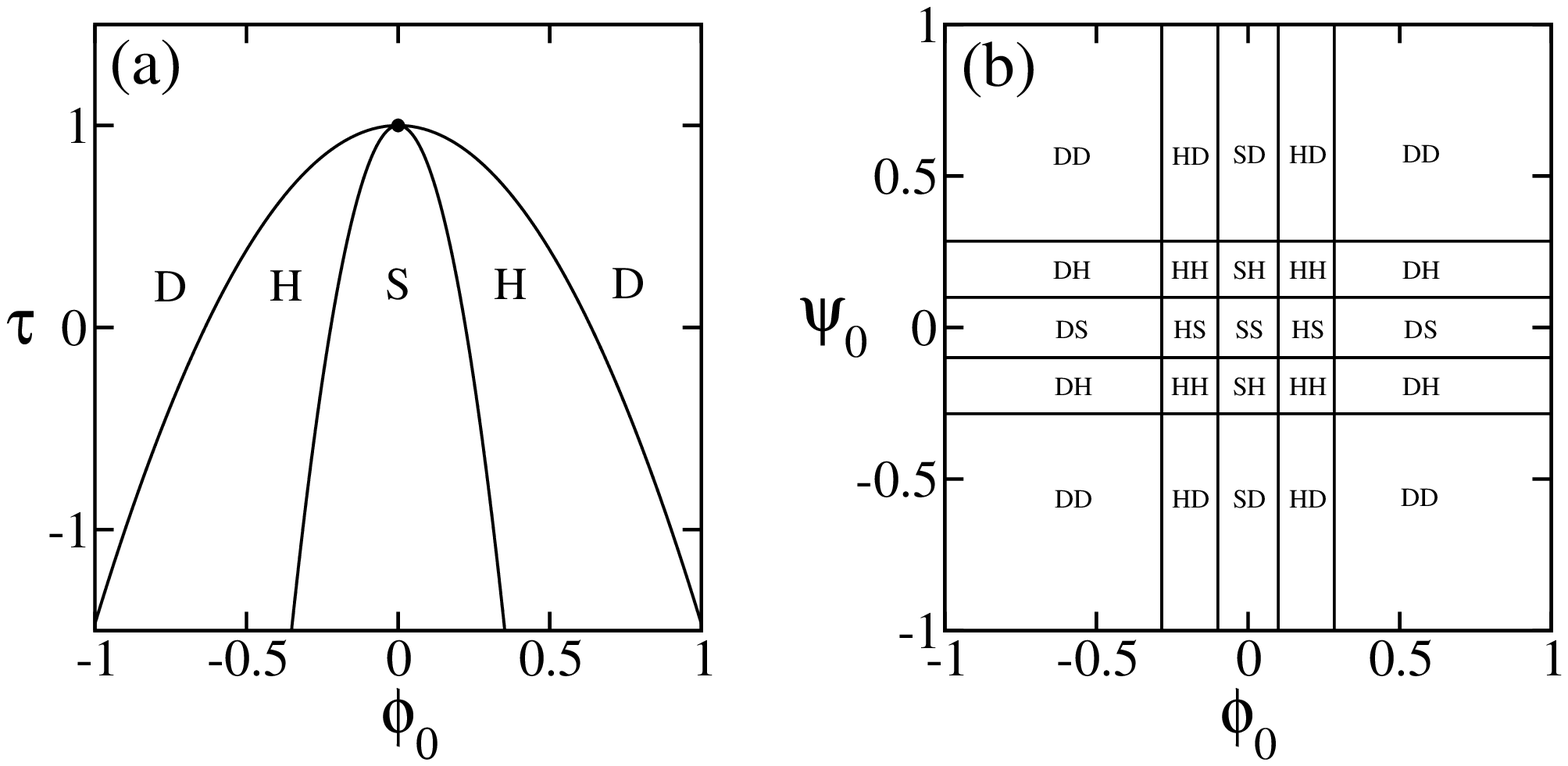}
\end{center}
\caption{\textsf{
(a) Mean-field phase diagram of a single monolayer with a modulated
structure in the vicinity of the critical temperature, computed
using a model as in Eqs.~(\ref{energyS})-(\ref{energyD}).
$\phi_0$ is the average composition and $\tau$ is the reduced
temperature.
The three phases are: striped (S),
hexagonal (H), and disordered (D).
These phases are separated by first-order transition lines, while for
clarity we omit showing coexistence regions.
The filled circle located at $(\phi_0,\tau)=(0,1)$ indicates
the critical point.
Note the shift of the critical temperature from zero to unity when the
modulated phases are considered.
(b) Mean-field phase diagram of decoupled ($\Lambda=0$) modulated
monolayers at $\tau=0.8$.
$\phi_0$ and $\psi_0$ are the average compositions in the two leaflets.
The notations of the different phases are described in the text
(see Sec.~\ref{density}).
All the phases are separated by first-order transition lines.
}}
\label{fig2}
\end{figure}

The phase behavior for uncoupled case, $\Lambda=0$, can be obtained 
from the analysis of $F_{\rm u}[\phi]$~\cite{LA} and is only briefly
reviewed here (see also Fig.~\ref{fig2}).
For a 2D system, the mean-field phase diagram can be constructed by
comparing the free energies of striped (S) and hexagonal (H) phases.
In terms of the  $\phi$ order parameter, the stripe phase is described by
\begin{equation}
\phi_{\rm S}(\mathbf{r})=\phi_0 + 2 \phi_q \cos(q^{\ast}x),
\end{equation}
where $\phi_0=\langle \phi \rangle$ is the spatially averaged
composition (imposed by the chemical potential $\mu_{\phi}$),
and $\phi_q$ is the amplitude of the $q^{\ast}$-mode in the
$x$-direction.
Similarly, the composition of the hexagonal phase is given
by a superposition of three 2D modes of equal magnitude, 
$\vert \mathbf{q}_i \vert=q^\ast$
\begin{equation}
\phi_{\rm H}(\mathbf{r}) = \phi_0 +
\frac{2\phi_q}{\sqrt{3}}\sum_{i=1}^3\,\cos(\mathbf{q}_i\cdot\mathbf{r}),
\end{equation}
where
\begin{eqnarray}
\mathbf{q}_1&=&q^\ast\hat{x}, 
\nonumber \\
\mathbf{q}_2&=&\frac{q^\ast}{2}\left(-\hat{x}+\sqrt{3}\hat{y}\right), 
\nonumber \\
\mathbf{q}_3&=&\frac{q^\ast}{2}\left(-\hat{x}-\sqrt{3}\hat{y}\right), 
\label{three_q}
\end{eqnarray}
and $\sum_{i=1}^3\mathbf{q}_i=0$.
In the above, only the most unstable wavenumber $q^{\ast}$ is
used within the single-mode approximation.
This can be justified for the weak segregation region close to
the critical point~\cite{GD}.

Averaging over one spatial period, we obtain the free energy
densities of the striped, hexagonal, and disordered phases, respectively
\begin{equation}
f_{\rm S} (\phi_0, \phi_q)=
\frac{\tau}{2} \phi_0^2 + \frac{1}{4} \phi_0^4
+ (\tau-1+3\phi_0^2)\phi_q^2 + \frac{3}{2}\phi_q^4,
\label{energyS}
\end{equation}
\begin{equation}
f_{\rm H}(\phi_0, \phi_q) =
\frac{\tau}{2} \phi_0^2 + \frac{1}{4} \phi_0^4
+ (\tau-1+3\phi_0^2)\phi_q^2
+ \frac{4}{\sqrt{3}}\phi_0 \phi_q^3
+ \frac{5}{2}\phi_q^4,
\label{energyH}
\end{equation}
\begin{equation}
f_{\rm D}(\phi_0) =
\frac{\tau}{2} \phi_0^2 + \frac{1}{4} \phi_0^4,
\label{energyD}
\end{equation}

In Fig.~\ref{fig2}(a), we reproduce the original phase diagram
of Ref.~\cite{ABJ,LA}.
The striped, hexagonal, and disordered phases are separated by
first-order phase-transition lines.
Regions of two-phase coexistence do exist but are omitted from
the figure for clarity sake~\cite{YK}.
Thus, the transition lines indicate the locus of points at which
the free energies of two different phases cross each other, and are not
the proper phase boundaries (binodals).
The critical point (filled circle) is located at
$(\phi_0,\tau)=(0,1)$.

\section{Two Coupled Leaflets with the Same  $q^\ast$}
\label{identical}

Having introduced the free energy and explained the phase behavior
of the uncoupled case, we shall now explore the equilibrium and
non-equilibrium properties of two coupled modulated monolayers,
$\Lambda \neq 0$.

\subsection{Free Energy Densities}
\label{density}

First we consider the case when $D=C=1$ so that the preferred
wavenumbers are the same for both monolayers,
$q_{\phi}^{\ast}=q_{\psi}^{\ast}=q^{\ast}=1/\sqrt{2}$.
The mean-field phase diagram is calculated within the single-mode
approximation.
Various combinations of 2D modulated structures appearing in the
two monolayers are possible.
The first example is the striped-striped (SS) phase, in which both
monolayers exhibit the striped phase.
This can be expressed as
\begin{equation}
\phi_{\rm S}(\mathbf{r})=\phi_0 + 2 \phi_q \cos(q^{\ast}x),
\end{equation}
\begin{equation}
\psi_{\rm S}(\mathbf{r})=\psi_0 + 2 \psi_q \cos(q^{\ast}x),
\end{equation}
where $\phi_0=\langle \phi \rangle$ and $\psi_0=\langle \psi \rangle$
are the average compositions, $\phi_q$ and $\psi_q$
are the respective amplitudes.
These composition profiles are substituted into the free energy
of Eq.~(\ref{freeenergy}).
Averaging over one spatial period, we obtain the free energy
density of the SS phase:
\begin{equation}
f_{\rm SS} = f_{\rm S}(\phi_0,\phi_q)
+ f_{\rm S}(\psi_0,\psi_q)
- \Lambda (\phi_0 \psi_0 + 2 \phi_q \psi_q),
\label{energySS}
\end{equation}
where $f_{\rm S}$ is defined in Eq.~(\ref{energyS}).
We then minimize $f_{\rm SS}$ with respect to both $\phi_q$ and
$\psi_q$ for given $\phi_0$, $\psi_0$, $\tau$ and  $\Lambda$.
When either $\phi_q$ or $\psi_q$ vanishes, the corresponding
monolayer is in its disordered phase and the  mixed bilayer
state will be called the striped-disordered (SD) or the
disordered-striped (DS) phase.
Note that we use the convention that the first index is of the
$\phi$--leaflet and the second of the $\psi$--one.
When both $\phi_q$ and $\psi_q$ are zero,
the free energy density of the disordered-disordered (DD) phase
is given by
\begin{equation}
f_{\rm DD} = f_{\rm D}(\phi_0) + f_{\rm D}(\psi_0)
- \Lambda \phi_0 \psi_0,
\label{energyDD}
\end{equation}
where $f_{\rm D}$ is defined in Eq.~(\ref{energyD}).
This free energy $f_{\rm DD}$ was analyzed in Ref.~\cite{PS} in order
to investigate the macro--phase separation of a bilayer membrane with
coupled monolayers.
It was shown that the bilayer can exist in four different phases, and
can also exhibit a three-phase coexistence.

Similar to the stripe case, the order parameters of the
hexagonal-hexagonal (HH) phase can be represented as
\begin{equation}
\phi_{\rm H}(\mathbf{r}) = \phi_0 +
\frac{2\phi_q}{\sqrt{3}}
\sum_{i=1}^3\,\cos(\mathbf{q}_i\cdot\mathbf{r}),
\label{HHphi}
\end{equation}
\begin{equation}
\psi_{\rm H}(\mathbf{r}) = \psi_0 +
\frac{2\psi_q}{\sqrt{3}} \sum_{i=1}^3\,\cos(\mathbf{q}_i\cdot\mathbf{r}).
\label{HHpsi}
\end{equation}
Where the basis of the three $\mathbf{q}_i$ was defined in Eq.~(\ref{three_q}).
By repeating the same procedure as for the SS phase, the free energy
density of the HH phase is obtained as
\begin{equation}
f_{\rm HH} = f_{\rm H}(\phi_0,\phi_q)
+ f_{\rm H}(\psi_0,\psi_q)
- \Lambda (\phi_0 \psi_0 + 2 \phi_q \psi_q),
\label{energyHH}
\end{equation}
where $f_{\rm H}$ is defined in Eq.~(\ref{energyH}).
When either $\phi_q$ or $\psi_q$ vanishes, one of the monolayers
is in the disordered phase and the bilayer will be called the
hexagonal-disordered (HD) phase or the disordered-hexagonal (DH)
phase.

When the normal hexagonal phase in one leaflet is coupled to the
inverted hexagonal phase in the other leaflet, it is energetically
favorable to have a particular phase shift of $2\pi/3$
between the two hexagonal structures.
The order parameters which represent such a different type of
hexagonal-hexagonal (HH$^{\ast}$) phase can be written as
\begin{equation}
\phi_{\rm H}(\mathbf{r}) = \phi_0 +
\frac{2\phi_q}{\sqrt{3}} \sum_{i=1}^3\,\cos(\mathbf{q}_i\cdot\mathbf{r}),
\label{HH'phi}
\end{equation}
\begin{equation}
\psi_{\rm H^{\ast}}(\mathbf{r}) = \psi_0 +
\frac{2\psi_q}{\sqrt{3}}
\sum_{i=1}^3\,\cos(\mathbf{q}_i\cdot\mathbf{r}+\frac{2\pi}{3}).
\label{HH'psi}
\end{equation}
The free energy density of the HH$^{\ast}$ phase is then obtained as
\begin{equation}
f_{\rm HH^{\ast}} = f_{\rm H}(\phi_0,\phi_q)
+ f_{\rm H}(\psi_0,\psi_q)
- \Lambda (\phi_0 \psi_0 - \phi_q \psi_q).
\label{energyHH'}
\end{equation}

Another combination which should be considered in the present
model is the asymmetric case where one monolayer exhibits the
striped phase and the other the hexagonal phase.
This striped-hexagonal (SH) phase is expressed as
\begin{equation}
\phi_{\rm S}(\mathbf{r}) =\phi_0 + 2 \phi_q \cos(q^{\ast}x),
\label{SHphi}
\end{equation}
\begin{equation}
\psi_{\rm H}(\mathbf{r}) = \psi_0 +
\frac{2\psi_q}{\sqrt{3}} \sum_{i=1}^3\,\cos(\mathbf{q}_i\cdot\mathbf{r}).
\label{SHpsi}
\end{equation}
The free energy density of this SH phase is calculated to be
\begin{equation}
f_{\rm SH} = f_{\rm S}(\phi_0,\phi_q)
+f_{\rm H}(\psi_0,\psi_q)
- \Lambda \left( \phi_0 \psi_0 +
\frac{2}{\sqrt{3}} \phi_q \psi_q \right).
\label{energySH}
\end{equation}
The phase in which $\phi_{\rm S}$ and $\psi_{\rm H}$
in Eqs.~(\ref{SHphi}) and (\ref{SHpsi}) are interchanged with
$\phi_{\rm H}$ and $\psi_{\rm S}$ is called the
hexagonal-striped (HS) phase, and its free energy is obtained from the
SH phase by noting the $\phi \leftrightarrow \psi$ symmetry.
In addition to these phases, we have also taken into account
the square-square (QQ) phase expressed by
\begin{equation}
\phi_{\rm Q}(\mathbf{r}) =\phi_0 +
\frac{2\phi_q}{\sqrt{2}} \left[
\cos(q^{\ast}x) + \cos(q^{\ast}y) \right],
\label{QQphi}
\end{equation}
\begin{equation}
\psi_{\rm Q}(\mathbf{r}) = \psi_0 +
\frac{2\psi_q}{\sqrt{2}} \left[
\cos(q^{\ast}x) + \cos(q^{\ast} y) \right].
\label{QQpsi}
\end{equation}
Then its free energy density is given by
\begin{equation}
f_{\rm QQ} = f_{\rm Q}(\phi_0,\phi_q)
+ f_{\rm Q}(\psi_0,\psi_q)
- \Lambda \left( \phi_0 \psi_0 +
2 \phi_q \psi_q \right),
\label{energyQQ}
\end{equation}
where
\begin{equation}
f_{\rm Q}(\phi_0,\phi_q)=
\frac{\tau}{2} \phi_0^2 + \frac{1}{4} \phi_0^4
+ (\tau-1+3\phi_0^2)\phi_q^2
+ \frac{9}{4}\phi_q^4.
\label{energyQ}
\end{equation}
However, we will show below that this QQ phase cannot be more
stable than the other phases.

\subsection{Bilayer Phase Diagrams}

\begin{figure}[t!]
\begin{center}
\includegraphics[scale=0.8]{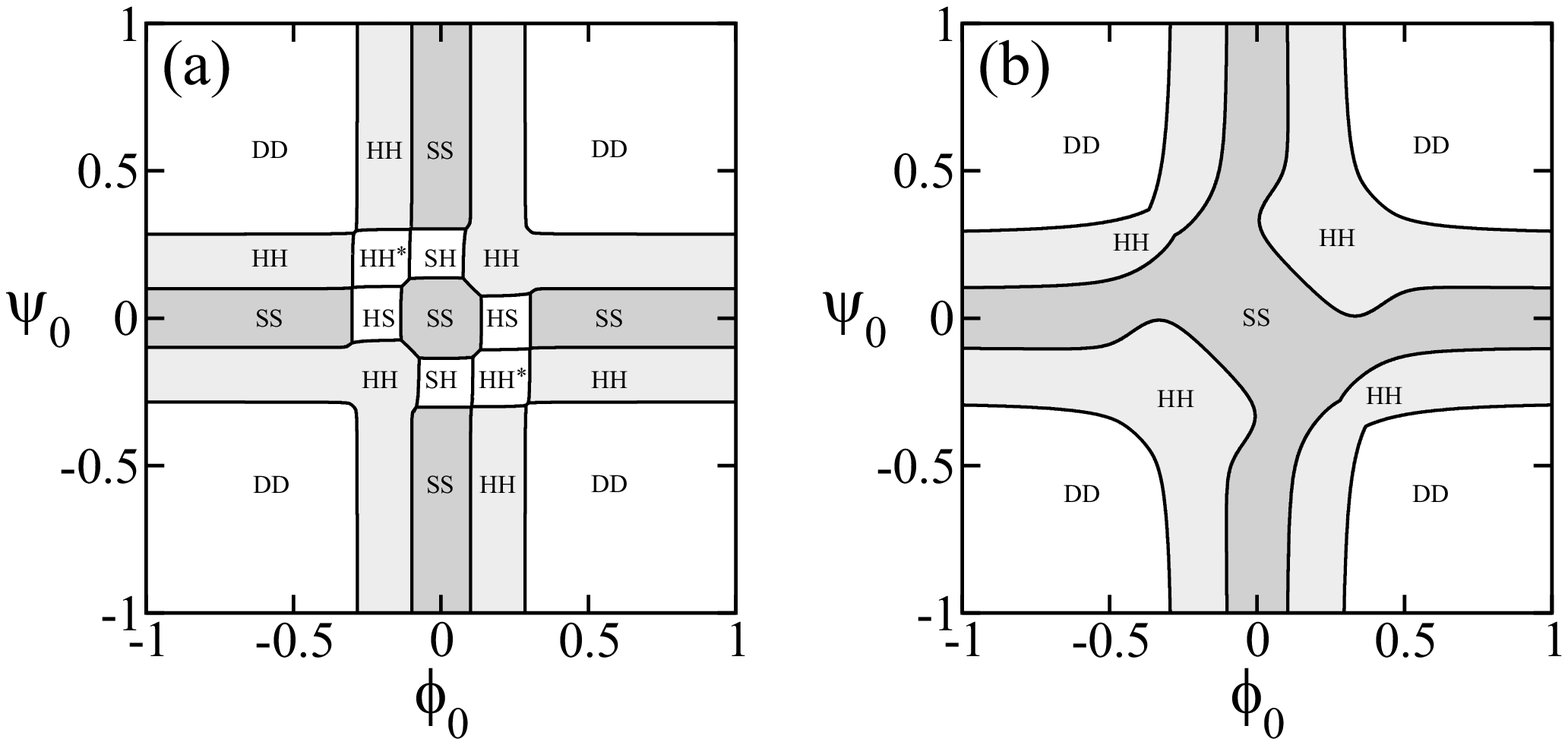}
\end{center}
\caption{\textsf{
Mean-field phase diagram of coupled modulated bilayers
for $\tau=0.8$.
$\phi_0$ and $\psi_0$ are the average compositions in the two
leaflets.
The coupling parameter is chosen to be (a) $\Lambda=0.02$ and
(b) $\Lambda=0.2$.
The notations of the different phases are described in the text
(see Sec.~\ref{density}).
All the phases are separated by first-order transition lines.
The phase diagram is symmetric with respect to the two principal 
diagonals $\phi_0=\psi_0$ and $\phi_0=-\psi_0$, as described in the
text.
}}
\label{fig3}
\end{figure}

Minimizing Eqs.~(\ref{energySS}), (\ref{energyHH}), (\ref{energyHH'}),
(\ref{energySH}) and (\ref{energyQQ}) with respect to both $\phi_q$
and $\psi_q$, we obtain the phase diagram for the coupled bilayer.
As a reference, we first show in Fig.~\ref{fig2}(b) the phase diagram
in the decoupled case ($\Lambda=0$) for $\tau=0.8$.
This can easily be obtained from Fig.~\ref{fig2}(a) by combining its
two cross-sections (one for $\phi_0$ and one for $\psi_0$) at
$\tau=0.8$.
Figure~\ref{fig3} gives the phase diagram for a coupled bilayer
when (a) $\Lambda=0.02$ and (b) $\Lambda=0.2$, while the temperature
is fixed to $\tau=0.8$ as before.
On the $(\phi_0,\psi_0)$-plane, we have identified the phase
which has the lowest energy, whereas possible phase coexistence regions
between different phases have been ignored.
All the boundary lines indicate first-order transitions.
Since the free energy Eq.~(\ref{freeenergy}) is invariant under
the exchange of $\phi \leftrightarrow \psi$, the phase diagrams
are symmetric about the diagonal line $\phi_0=\psi_0$
as the upper and lower leaflets have been chosen arbitrarily.
These phase diagrams are also symmetric under the rotation of 
180 degrees around the origin because Eq.~(\ref{freeenergy}) is
invariant (except the linear terms) under the simultaneous
transformations of $\phi \rightarrow - \phi$ and
$\psi \rightarrow -\psi$.
This is reasonable as the labels of ``A'' or ``B'' for the
two lipids have been assigned arbitrarily. 
As a consequence, the phase diagrams are also symmetric 
about the diagonal line $\phi_0=-\psi_0$. 
The symmetries with respect to both $\phi_0=0$ and $\psi_0=0$ in
Fig.~\ref{fig2}(b) for $\Lambda=0$ are now broken because of the
coupling between the two leaflets.

When the coupling parameter is small ($\Lambda=0.02$), the
global topology of the phase diagram resembles that of
the uncoupled case presented in Fig.~\ref{fig2}(b).
Close to the origin, $\phi_0=\psi_0=0$, there is a region of
SS phase surrounded by eight other phases: two SH, two HS, two HH,
and two HH$^{\ast}$ phases.
The HH phase appearing in the region of $\phi_0<0$ and $\psi_0<0$
is the combination of the two inverted hexagonal structures on each
monolayer.
One sees that the HH$^{\ast}$ phase appears in the regions of
$\phi_0 \psi_0<0$, where the hexagonal and the inverted hexagonal
structures are coupled to each other.

A remarkable feature of this phase diagram is the
existence of the SS and HH phases in the regions where
either $\vert \phi_0 \vert$ or $\vert \psi_0 \vert$ are large.
These outer SS and HH phases extend up to the maximum
or the minimum values of the compositions.
These regions of the SS and HH phases with $\Lambda > 0$
roughly correspond to those of the SD (DS) and HD (DH) phases,
respectively, in Fig.~\ref{fig2}(b) with $\Lambda = 0$.
Hence the modulated structure in one of the monolayers induces the
same modulated phase in the other monolayer due to the coupling term.
Notice that the SD (DS) phase and HD (DH) phase do not exist in
Fig.~\ref{fig3}(a).
We further remark that the extent of the four DD phase regions is
almost unaffected by the coupling.
Even when the temperature is lowered by decreasing $\tau$, only the
phases located close to the origin ($\phi_0=\psi_0=0$) would expand,
and the global topology does not change substantially.

For a larger value of the coupling parameter ($\Lambda=0.2$),
the five regions of the SS phase merge together 
forming one single continuous SS region.
The four HH regions and still distinct and
separate the SS region from four DD phase regions.
Note that in Fig.~\ref{fig3}(b), all phases have a symmetric
combination of phase modulation such as SS or HH.
The asymmetric combination such as the SH phase does not appear,
because the large coupling parameter strongly
prefers symmetric phases of equal modulations in the two monolayers, although 
the $\phi_q$ and $\psi_q$ amplitudes of the two
modulated monolayers are not the same in the 
stripe SS phase (or the hexagonal HH phase).
As the value of $\Lambda$ is increased from $0.02$ to $0.2$,
first the SH phase disappears, followed by the disappearance of 
the HH$^{\ast}$ phase.
When the value of $\Lambda$ is further increased, the regions of
the SS and HH phases expand on the expense of the DD phase regions.
This means that the coupling between the monolayers causes
more structural order in the bilayer.
Finally we remark that the QQ phase
was never found to be more stable than any of the other phases considered
above.

Although we have so far assumed that $\Lambda$ is positive, the
phase diagrams for $\Lambda<0$ can be easily obtained from those
for $\Lambda>0$ by rotating them by 90 degrees around the origin.
This is because the free energy Eq.~(\ref{freeenergy}) is invariant
under the simultaneous transformations of either
$\phi \rightarrow - \phi$ and $\Lambda \rightarrow - \Lambda$, or
$\psi \rightarrow - \psi$ and $\Lambda \rightarrow - \Lambda$.

\subsection{Modulated Bilayer Dynamics}

\begin{figure}[t!]
\begin{center}
\includegraphics[scale=0.75]{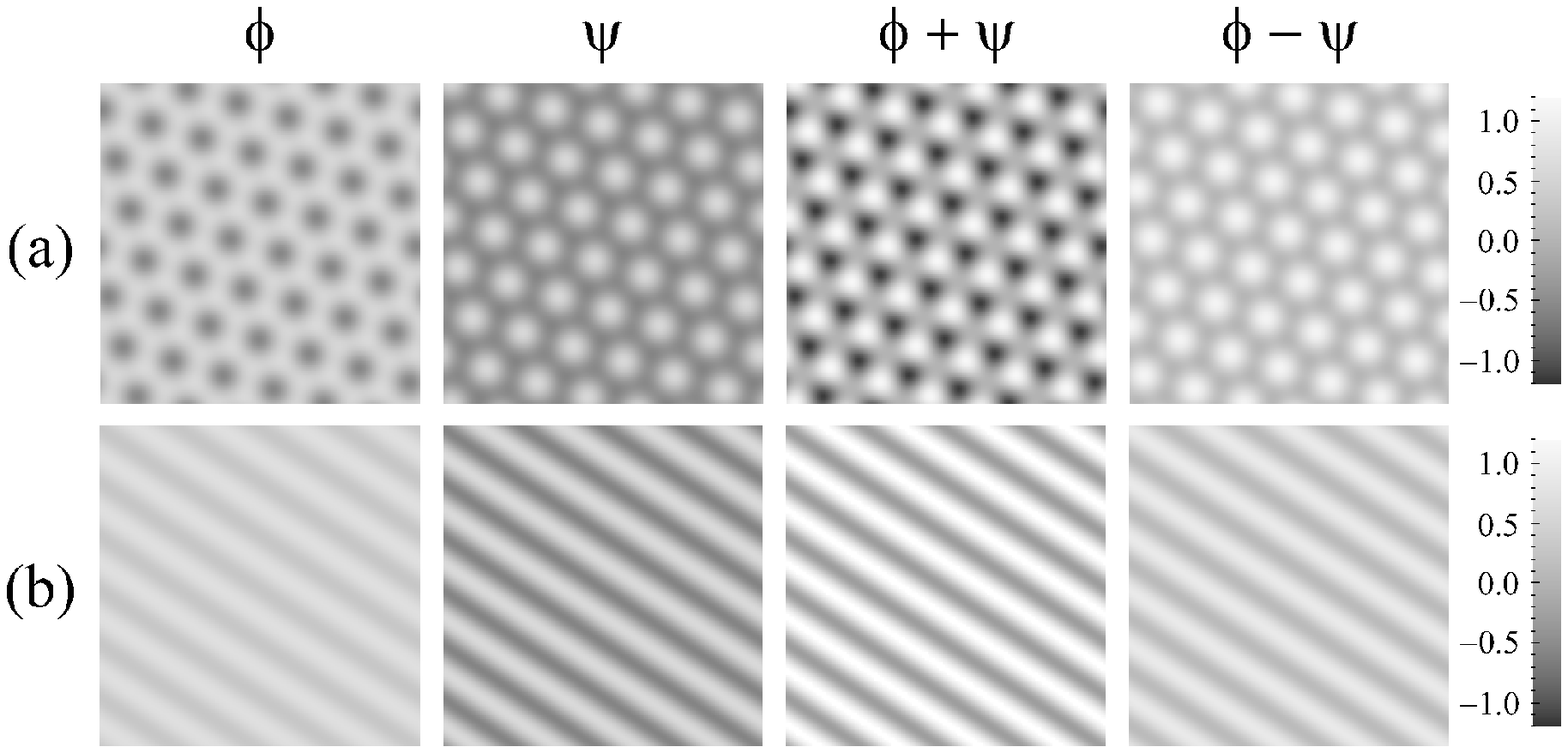}
\end{center}
\caption{\textsf{
Equilibrium patterns of coupled modulated monolayers with $\tau=0.8$. In addition
setting $D=C=1$ implies the same $q^\ast$ in both monolayers.
The patterns of $\phi$, $\psi$, $\phi+\psi$, and $\phi-\psi$
at $t=5,000$ are presented.
The other parameters are chosen to be
(a) $\phi_0=0.2$, $\psi_0=-0.2$, $\Lambda=0.02$, and
(b) $\phi_0=0.5$, $\psi_0=0$, $\Lambda=0.2$.
}}
\label{fig4}
\end{figure}

In order to check the validity of the obtained phase diagram
and to investigate the dynamics of coupled modulated bilayers,
we consider now the time evolution of the coupled equations for 
$\phi$ and $\psi$:
\begin{equation}
\frac{\partial \phi}{\partial t}=
L_{\phi} \nabla^2 \frac{\delta F}{\delta \phi},~~~~~~
\frac{\partial \psi}{\partial t}=
L_{\psi} \nabla^2 \frac{\delta F}{\delta \psi}.
\label{evolution}
\end{equation}
Here we have assumed that both $\phi$ and $\psi$ are conserved
order parameters in each of the monolayer (model B in the
Hohenberg-Halperin classification~\cite{HH}).
For simplicity, the kinetic coefficients $L_{\phi}$ and $L_{\psi}$
are taken to be unity, and both the hydrodynamic effect and thermal
fluctuations are neglected.
We solve the above equations numerically in 2D using the periodic
boundary condition.
Each simulation starts from a disordered state with a small random
noise around the average compositions $\phi_0$ and $\psi_0$.
In Fig.~\ref{fig4}, we show typical equilibrium  patterns
of $\phi$, $\psi$, $\phi+\psi$ and $\phi-\psi$ for two choices of
parameters.
The $\phi+\psi$ pattern is presented here because this quantity
can be directly observed in the experiment on Montal-Mueller bilayers
using fluorescence microscopy~\cite{CK}.
The quantity $\phi-\psi$ measures the concentration contrast between the $\phi$
and $\psi$ leaflets.
Time is measured in discrete time steps, and $t=5,000$ corresponds
to a well equilibrated system.
In all the simulations below, the temperature is fixed to be
$\tau=0.8$ corresponding to the weak segregation regime.
Notice that all the patterns in Fig.~\ref{fig4} are presented
with the same gray scale.

Figure~\ref{fig4}(a) illustrates the coupling between a hexagonal
phase with $\phi_0=0.2$ and an inverted hexagonal phase with
$\psi_0=-0.2$ in the weak coupling regime ($\Lambda=0.02$).
Being consistent with the phase diagram of
Fig.~\ref{fig3}(a), this parameter choice yields the HH$^{\ast}$ phase as seen
from the pattern of $\phi+\psi$ where the two hexagonal structures
are superimposed.
We note that the difference in the order parameter $\phi-\psi$ also
exhibits a hexagonal structure.

Figure~\ref{fig4}(b) shows the equilibrium patterns for
$\phi_0=0.5$ and $\psi_0=0$ in the strong coupling regime ($\Lambda=0.2$).
If there were no coupling, the $\phi$-monolayer would not exhibit
any modulation (as it is in its own disordered phase), whereas the
$\psi$-monolayer is in the striped phase.
We clearly see that, due to the coupling effect, the stripe structure
is induced in the pattern of $\phi$.
This corresponds to the SS phase shown in Fig.~\ref{fig3}(b).
The periodicities of the two striped structures are the same, although
their amplitudes differ.
Notice that the modulation phase of $\phi+\psi$ is 
shifted by $\pi$ relatively to that of $\phi-\psi$.
Since the patterns in Fig.~\ref{fig4} would correspond to the
equilibrium configurations, they can be compared with the phase
diagrams in Fig.~\ref{fig3}.
We conclude that these simulation results indeed reproduce the
predicted equilibrium modulated structures.

\section{Coupled monolayers with two different $q^\ast$}
\label{different}

We consider next the more general case in which the preferred
wavelengths of modulation in the two uncoupled leaflets are different,
$q^\ast_\phi \neq q^\ast_\psi$.
The free energy densities cannot be obtained analytically
as was done in Sec.~\ref{density}, because there is not a single
periodicity on which one can average $\phi(\mathbf{r})$ and
$\psi(\mathbf{r})$.
Due to such a difficulty in the analytical treatment, we 
present below the results of numerical simulations, relying on
Eq.~(\ref{evolution}) for the time evolution of the
two coupled order parameters.

\begin{figure}[t!]
\begin{center}
\includegraphics[scale=0.75]{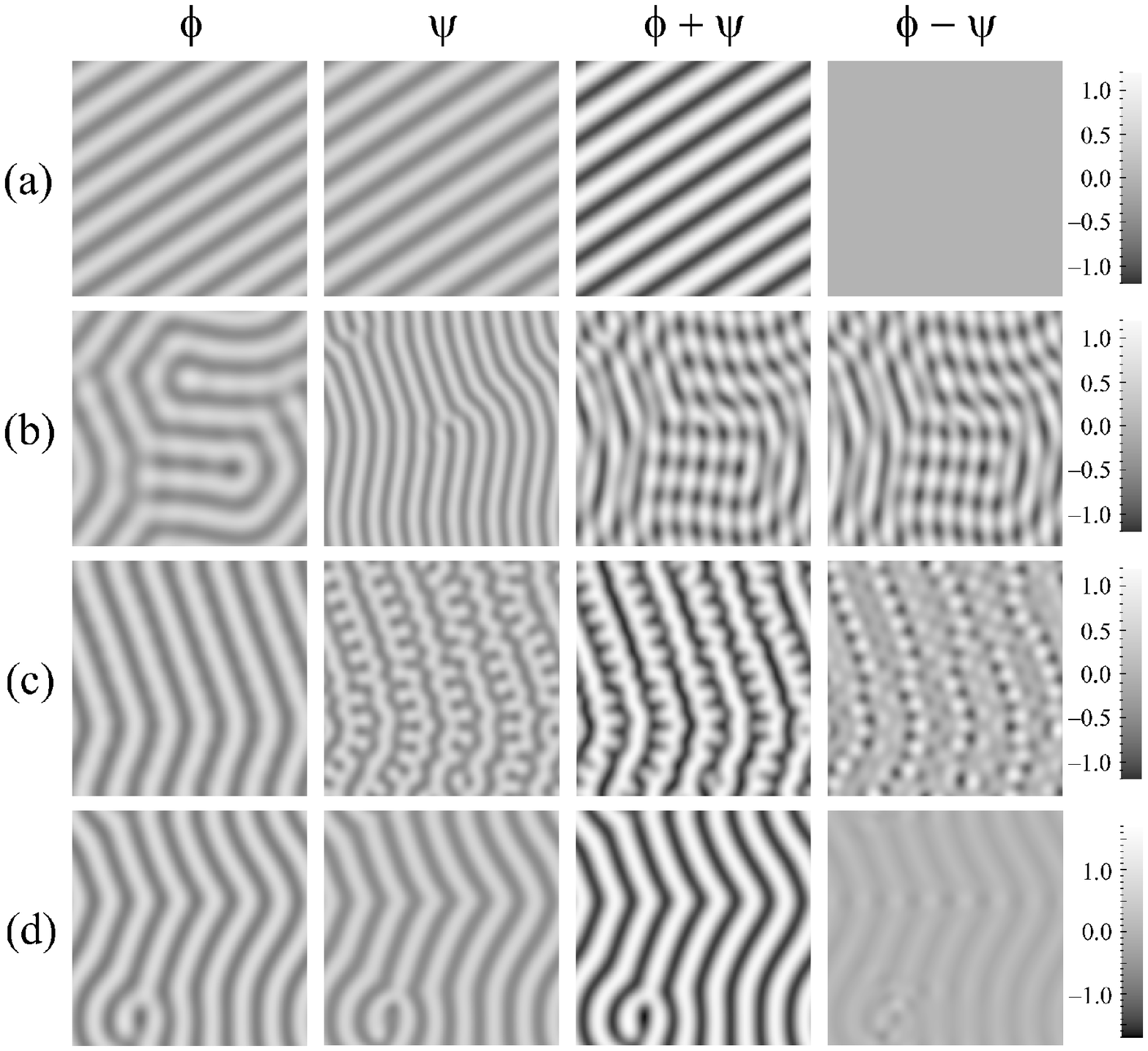}
\end{center}
\caption{\textsf{
Patterns of coupled modulated monolayers
with $\tau=0.8$.
The patterns of $\phi$, $\psi$, $\phi+\psi$, and $\phi-\psi$
are presented for simulation time, $t=5,000$, and 
the average compositions are
set to be $\phi_0=\psi_0=0$.
The other parameters are chosen to be
(a) $D=C=1$, $\Lambda=0.02$,
(b) $D=0.1296$, $C=0.36$, $\Lambda=0.02$,
(c) $D=0.1296$, $C=0.36$, $\Lambda=0.2$, and
(d) $D=0.1296$, $C=0.36$, $\Lambda=0.4$. In all cases but (a), $D\neq C$ and the 
two periodicities are nonequal, $q^\ast_\phi\neq q^\ast_\psi$.
}}
\label{fig5}
\end{figure}

\begin{figure}[t!]
\begin{center}
\includegraphics[scale=0.8]{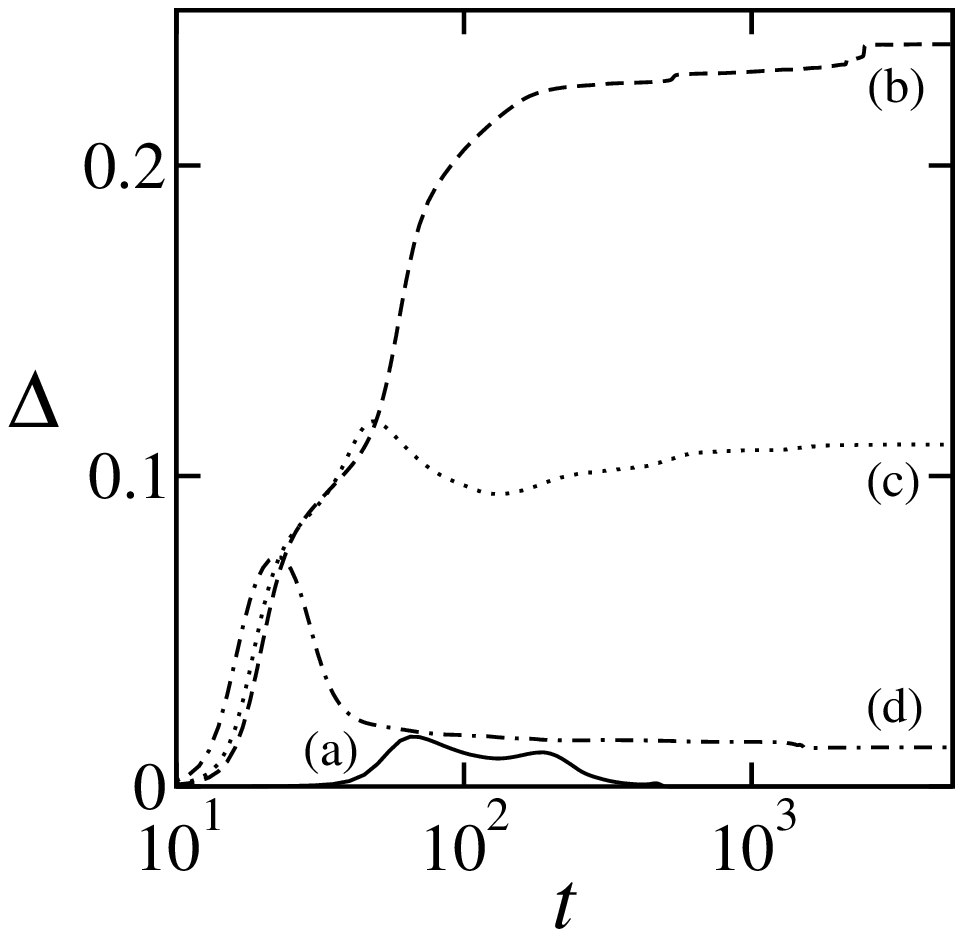}
\end{center}
\caption{\textsf{
Time evolutions of $\Delta$ defined by Eq.~(\ref{delta}).
The solid, dashed, dotted, and dashed-dotted lines correspond to the
parameters of Fig.~\ref{fig5}(a), (b), (c) and (d), respectively.
}}
\label{fig6}
\end{figure}

In Fig.~\ref{fig5}, we show the patterns for $\tau=0.8$
and $\phi_0=\psi_0=0$, when both monolayers exhibit the
striped phase without the coupling.
As a reference, we show in Fig.~\ref{fig5}(a) the case when $D=C=1$
and $\Lambda=0.02$ corresponding to the SS phase in Fig.~\ref{fig3}(a).
The patterns of $\phi$ and $\psi$ match each other as the composition
difference $\phi-\psi$ vanishes throughout the system.
In Fig.~\ref{fig5}(b), the parameters are chosen to be $D=0.1296$,
$C=0.36$ and $\Lambda=0.02$.
The preferred wavenumbers of the two monolayers are different:
$q_{\psi}^{\ast} = 1.67q_{\phi}^{\ast}$ for uncoupled leaflets.
The above set of parameters, especially $D$ and $C$, are chosen in such
a way that the amplitudes of the two stripes are nearly equal.
As long as the coupling parameter is small, the two stripes of
different periodicities are formed rather independently.
The superposition of the two striped structures produces an interference
pattern resulting in a new modulation as seen from the pattern of $\phi+\psi$.
The striped modulations of $\phi$ and $\psi$ are almost parallel or
perpendicular to each other.

When $\Lambda$ is made larger, up to $\Lambda=0.2$ as
in Fig.~\ref{fig5}(c), the $\psi$ field exhibits a complex
pattern in which two different length scales coexist (reflecting
$q_{\phi}^{\ast}$ and $q_{\psi}^{\ast}$), whereas the
pattern of $\phi$ is characterized by a single mode
(reflecting $q_{\phi}^{\ast}$).
The patters of $\phi$ and $\psi$ almost match each other
when $\Lambda=0.4$ as seen in Fig.~\ref{fig5}(d).
In this case, the modulation with a longer wavelength
($q_{\phi}^{\ast}$) dominates both monolayers.
Figure \ref{fig5}(b), (c), (d) provide a typical sequence of
morphological changes, i.e., interference pattern $\rightarrow$
two-mode pattern $\rightarrow$ single-mode pattern, as the coupling
constant $\Lambda$ is increased.

To further analyze the temporal correlations of the
two order parameters, $\phi$ and $\psi$, we have plotted
in Fig.~\ref{fig6} the time evolution of the quantity
\begin{equation}
\Delta(t) = \frac{1}{L^2} \int {\rm d}\mathbf{r} \,
[\phi(\mathbf{r},t) - \psi(\mathbf{r},t)]^2,
\label{delta}
\end{equation}
where $L=128$ is the linear system size in the
simulations.
The solid, dashed, dotted, and dashed-dotted lines correspond to the
time evolutions of $\Delta(t)$ in Fig.~\ref{fig5}(a), (b), (c), and (d)
respectively.
The solid line (a) first increases and then approaches to zero since
the patterns of the two stripes coincide in the late stage.
The dashed line (b) increases in two separate stages.
At first because the growth rate of the modulation having smaller
wavelength (corresponding to $\psi$) is faster than that having
larger periodicity (corresponding to $\phi$) as is also revealed
from the linear stability analysis of Eq.~(\ref{evolution}) which 
will be published elsewhere.
At late temporal stages, the value of $\Delta$ remains large for small
coupling parameter ($\Lambda=0.02$).
When the coupling becomes even stronger ($\Lambda=0.2$) as for the dotted
line (c), the value of $\Delta$ is suppressed compared to the dashed
line (b), because $\phi$ and $\psi$ tend to have more overlap
for larger $\Lambda$.
The same applies for the dashed-dotted line (d) with $\Lambda=0.4$
as compared to the dotted line (c).

\section{Discussion}
\label{discussion}

We propose a minimal model describing
the coupling phenomena between two modulated bilayers.
Considering 2D case, we obtain the mean-field phase diagram when
the two coupled and spatially modulated monolayers have the same
preferred periodicity.
Various combinations of modulated phases can exist such as the
SS, HH, HH$^{\ast}$ and SH (HS) phases as described in
Sec.~\ref{identical}.
We have seen that modulations in one of the monolayers induces a
similar modulations in the other.
The region of the induced modulated phase expands as the coupling
parameter becomes larger.

When the two monolayers have different inherent wavelengths
in the decoupled case, we have conducted numerical simulations to
investigate the morphologies and dynamics of the coupled system.
We obtain several complex patterns arising from the frustration
induced by the two incommensurate structures.
As the coupling constant $\Lambda$ is made larger, the two different
modes start to interfere with each other and eventually coincide.
The time evolution of the striped structures can take place in two
steps reflecting the different growth rate of the two modulations.

It is instructive to rewrite the free energy Eq.~(\ref{freeenergy})
in terms of the sum and the difference of $\phi$ and $\psi$ 
with $\eta_\pm = \phi\pm \psi$.
When $D=C=1$, we obtain
\begin{align}
F[\eta_+, \eta_- ] & =\int {\rm d}\mathbf{r}
\biggl[(\nabla^2 \eta_+)^2 - (\nabla \eta_+)^2
+ \frac{1}{4}(\tau-\Lambda)\eta_+^2 + \frac{1}{32}\eta_+^4
-\mu_+ \eta_+
\nonumber \\
& + (\nabla^2 \eta_-)^2 - (\nabla \eta_-)^2
+ \frac{1}{4}(\tau+\Lambda)\eta_-^2 + \frac{1}{32}\eta_-^4
-\mu_- \eta_-
+\frac{3}{16}\eta_+^2 \eta_-^2 \biggr],
\label{freeenergy2}
\end{align}
where $\mu_\pm=(\mu_{\phi}\pm \mu_{\psi})/2$.
Hence the coupling term between $\eta_+$ and $\eta_-$ in the free 
energy takes the form of an $\eta_+^2 \eta_-^2$ term  with a numerical 
positive coefficient.
The original coupling parameter $\Lambda$ enters in the coefficients
of $\eta_+^2$ and $\eta_-^2$ terms (but not in the $\eta_+^2 \eta_-^2$ 
coupling term). 
It shifts the respective
transition temperatures of $\eta_+$ and $\eta_-$ in opposite directions.
When the gradient terms are absent and $\mu_-=0$, a similar model was
considered by MacKintosh and Safran who studied transitions between
lamellar and vesicle phases in two-component fluid bilayers~\cite{MS}.

The proposed free energy Eq.~(\ref{freeenergy}) has some analogies
to the previous model for the rippled phase in lipid bilayers~\cite{KK}.
It was argued that the coupling between the membrane curvature
and the asymmetry in the area per molecule between the two monolayers
would induce a structural modulation of a bilayer.
By considering a similar mechanism, Kumar \textit{et al.}~\cite{KGL} 
investigated various modulated phases in two-component bilayer membranes.
They claimed that the phase behavior of two-component bilayers
resembles that of three-component monolayer.
This is because the three different local combinations of upper/lower
composition in bilayers (A/B, B/A, and A/A for excess of A), would
correspond to three different types of molecules for the monolayer.
One of the new aspects in our model is that the preferred wavelengths
of the two monolayers can, in general, be different from one another leading
to a frustrated bilayer state.

We also point out that there are some similarities between coupled 
modulated structures and the problem of atoms adsorbed on a
periodic solid substrate.
The latter topic has been extensively studied within the
Frenkel-Kontorova (FK) model which provides a simple description of the
commensurate-incommensurate transition~\cite{CL}.
Our model and the FK model are analogous in the sense that there are
two natural length scales whose ratio changes as a function of other model
parameters.
In the FK model, however, these length scales are quenched,
whereas in our model they are annealed.

Another related experimental system can be seen for  surface-induced
ordering in thin film of diblock copolymers~\cite{TA}.
When the surface is periodically patterned, a tilt of the lamellae is
induced in order to match the surface periodicity.
The situation becomes more complex if a copolymer melt is confined
between two surfaces.
An interesting case arises when the spacing between the two surfaces
is incommensurate with the lamellar periodicity~\cite{TA}.

For systems out of equilibrium, spatial
resonances and superposition patterns combining stripes and/or
hexagons were investigated in a reaction-diffusion model with
interacting Turing modes of different wavelengths~\cite{YDZE}.
These models were successful in reproducing hexagonal superlattice
patterns which are known as ``black-eyes''.
Although the mechanism of pattern formation is different than
in our model, we observe similar superposition patterns
as reported in Ref.~\cite{YDZE}, such as hexagons on stripes or
hexagons on hexagons (not shown in this paper).

A more detailed study of the present model and several interesting
extensions will be published elsewhere.
One possible extension is to consider vector order parameters
describing, for example, the molecular tilt for coupled
bilayers~\cite{CLM,SSN}.
When the two order parameters are vectors, the nature of the transitions
between different phases can be different, and even the square phase
may exist in thermodynamic equilibrium~\cite{TLFKH}.

\begin{acknowledgments}

We thank T. Kato, S. L. Keller, M. Schick, and K. Yamada for useful
discussions.
YH acknowledges a Research Fellowship for Young Scientists
No.\ 215097 from the Japan Society for the Promotion of Science (JSPS).
SK acknowledges support by KAKENHI (Grant-in-Aid for Scientific
Research) on Priority Areas ``Soft Matter Physics'' and Grant
No.\ 21540420 from the Ministry of Education, Culture, Sports,
Science and Technology of Japan.
DA acknowledges support from the Israel Science Foundation (ISF)
under grant No.\ 231/08 and the US-Israel Binational Foundation (BSF)
under grant No.\ 2006/055.
\end{acknowledgments}


\end{document}